\documentstyle[epsf]{mn}
\input epsf.sty
\def\hmpc{~h$^{-1}$ Mpc~}

\title[A substructure analysis of the A3558 cluster complex]
 {A substructure analysis of the A3558 cluster complex}
\author[S. Bardelli et al.]{
S. Bardelli$^{1}$,
A. Pisani$^{1}$,
M. Ramella$^{1}$,
E. Zucca$^{2}$
\&
G. Zamorani$^{2,3}$
\\ $^1$ Osservatorio Astronomico di Trieste, 
via Tiepolo 11, I--34131 Trieste, Italy
\\ $^2$ Osservatorio Astronomico di Bologna, 
via Zamboni 33, I--40126 Bologna, Italy,
\\ $^3$ Istituto di Radioastronomia del CNR, via Gobetti 101, I-40129, Italy
\\ E-mail: bardelli@astrts.oat.ts.astro.it
}
\date{Received 00 - 00 - 0000; accepted 00 - 00 - 0000}
\begin{document}
\maketitle

\begin{abstract}
The ``algorithm driven by the density estimate for the identification of 
clusters" (DEDICA, Pisani 1993, 1996) is applied to the A3558 cluster 
complex in order to find substructures. This complex, located at the center
of the Shapley Concentration supercluster, is a chain formed by the ACO clusters
A3556, A3558 and A3562 and the two poor clusters SC 1327-312 and SC 1329-313.
We find a large number of clumps, indicating that strong dynamical processes
are active. In particular, it is necessary to use a fully 
three-dimensional sample
(i.e. using the galaxy velocity as third coordinate) in order to recover 
also the clumps superimposed along the line of sight. 
Even if a great number of detected substructures were already found in a 
previous analysis (Bardelli et al. 1998), this method is more efficient and 
faster when compared with the use of a wide battery of tests and permits
the direct estimate of the detection significance.
Almost all subclusters previously detected by the wavelet analyses found in 
the literature are recognized by DEDICA.

On the basis of the substructure analysis, we also briefly discuss the origin 
of the A3558 complex by comparing two hypotheses: 1) the structure is a 
cluster-cluster collision seen
just after the first core-core encounter; 2) this complex is the result of 
a series of incoherent group-group and cluster-group mergings, focused in 
that region by the presence of the surrounding supercluster.
We studied the fraction of blue galaxies in the detected substructures
 and found that the bluest groups reside between A3562 and A3558, i.e.
in the expected position in the scenario of the cluster-cluster collision.

\end{abstract}

\begin{keywords}
galaxies--
clusters--
substructures--
individuals: 
A3558--
A3562--
A3556--
SC 1329-313--
SC 1327-312--
SC 1329-314
\end{keywords}
%
%
%
\section{Introduction}
Clusters of galaxies are thought to form by accretion of subunits
in a hierarchical bottom up scenario. This view arises naturally in 
various theories of cosmic structure formation, as f.i. the Cold 
Dark Matter dominated scenarios. 
Numerical simulations on scales of cosmological relevance (e.g.
Cen \& Ostriker 1994, Colberg et al. 1997)  
revealed that mergings happen along preferential directions, called 
density caustics, which
define matter flows, at whose intersection rich clusters are formed.
Detailed high resolution simulations of such collisions were performed by 
McGlynn \& Fabian (1984) and Roettiger, Burns \& Loken (1993), who studied
cluster-cluster and cluster-group merging, respectively. 
In the simulations, the two cores have two or three encounters before 
dissipating all the
kinetic energy: in particular, the less massive cluster emerges as a 
partially dispersed galaxy population along the down merger side of the 
dominant cluster. Members of the merging cluster, trapped nearby the dominant 
core, create an apparent clumpiness on either side of the center. The kinetic 
energy is dissipated mostly by the particles in external regions of the 
clusters,  which expand creating a common envelope, while the cores do not 
change dramatically.

From the observational point of view, the detection of a large fraction
of clusters which present substructures (see f.i. Kriessler \& Beers 1997 and
references therein) revealed that these systems are still dynamically young
and show evidence of mergings.
The best studied examples are 
 A2256 
(Briel, Henry \& B\"ohringer 1992), where a small group is detected
in the X-ray band nearby the cluster center,
and Coma, where a number of substructures are 
revealed (Biviano et al. 1996). In particular, 
Burns et al. (1994), on the basis of the detection of a large scale
X-ray emitting filament connecting Coma with the NGC4839 group, suggest
that the group has probably already passed the cluster center.

The most spectacular phenomena are expected to be seen when the merging 
involves repeated collisions or even coalescence of comparable units: 
during the intermediate stages, the clusters form structures and
complexes on scales of few megaparsecs. 

Rich superclusters are the ideal environment for the detection of cluster 
mergings, because the peculiar velocities induced by the enhanced local density
of the large scale structure favour the cluster-cluster and cluster-group 
collisions, in the same way as the caustics seen in the simulations. 
The most remarkable examples of cluster merging seen at an intermediate stage
is found in the central region of the Shapley Concentration, the richest
supercluster of clusters found within $300$ \hmpc (Zucca et al. 1993;
hereafter h=H$_o$/100). On the basis
of the two-dimensional distribution of galaxies of the COSMOS/UKSTJ catalogue
(Yentis et al. 1992), it is possible to find several complexes of interacting
clusters, which suggest that the entire
region is dynamically active. Therefore, this supercluster represents a unique 
laboratory where it is possible to follow cluster mergings and to test related 
astrophysical consequences,
as the formation of radio halos, relicts and wide angle tail radio sources and 
the presence of post-starburst (E+A) galaxies.

In this paper we perform a substructure analysis of the A3558 cluster complex
using a non-parametric, scale independent algorithm, based on 
the adaptive kernel density estimator. This method could be considered
alternative or complementary to the wavelet decomposition analysis, extensively
used in literature. 
In particular, a wavelet analysis of this complex was done by Girardi et al. 
(1997) and, limited to the dominant cluster A3558, by
Dantas et al. (1997): these analyses offer an opportunity to test the two
methods on a real, rather complex situation. 
In Section 2 we briefly describe the method, in Section 3 the used samples
are presented and in Section 4 a brief discussion of the subclumps is reported;
finally in Section 5 we present our conclusions. 

\section{The method}

In order to identify the substructures in our sample and to assess their 
significance, we followed the method known as ``algorithm driven by the 
density estimate for the identification of clusters" (hereafter DEDICA) 
developed by Pisani (1993, 1996). 
This is a non-parametric, scale independent method, whose output gives a list 
of groups with the related statistical significance, their members, and a list 
of isolated objects.
Here we remind briefly the main steps of the method.

First, we estimated the density field $f(\vec{x})$ (where $\vec{x}$ is the
$n$-dimensional position vector) using the ``adaptive kernel probability 
estimator" for which
\begin{equation}
f_k(\vec{x})=  {1\over N} \sum^N_{i=1} K(\vec{x}_i,\sigma_i; \vec{x}) 
\end{equation}
where $f_k(\vec{x})$ is the kernel estimator of $f(\vec{x})$, $\vec{x}_i$ 
is the position of the $i^{th}$ galaxy, $N$ is the number of objects
in the sample and $\sigma_i$ is the size of the 
gaussian kernel defined as
\begin{equation}
K(\vec{x}_i, \sigma_i; \vec{x})={ 1 \over{ \left(2 \pi \sigma_i^2 \right)
^{d/2} }} exp \left( - { {|\vec{x}_i- \vec{x}|^2} \over{2 \sigma_i^2}}\right)
\end{equation}
where $d$ is the number of dimensions of the sample. 

We stress that we choose the Gaussian form for the kernel basically because
 it is differentiable in all its domain. This is a crucial feature
in order to identify the density peaks according to Eq. \ref{eq:path}.
Even if the most efficient (i.e. with minimum variance) kernel 
form is the Epanechnikov function, Silverman (1986) showed that difference
in the use of the two forms is negligible. Indeed, what is crucial
for cluster detection is the width and not the shape of the kernels,
provided that they satisfy the normalization and convergence condition
(see section I of Pisani 1993).

The estimate of the kernel sizes is done through an iterative procedure, 
starting from a large value for the kernel sizes and progressively reducing 
them until the minimum of the ``cross-validation function" is reached.
The various steps are the following (for details see Pisani 1993, 1996):
\par\noindent
1) set the first guess ($n=0$) of the size as $\sigma_0 = 4 \sigma_t$ where
\begin{equation}
\sigma_t= 0.96 N^{[-1/(d+4)]} \sqrt{ {1 \over d} \sum^{d}_{l=1} s_{l}^2}
\end{equation}
where $s_{l}^2$ is the standard deviation of the $l^{th}$ coordinate 
(Silverman 1986).
In this step, a large starting $\sigma_0$ is selected: note that for a symmetric
gaussian distribution with dispersion $\sigma_G$ and same number of points 
of our samples, we have $\sigma_0 \sim 1-1.5 \sigma_G$; 
\par\noindent
2) set the following iteration ($n^{th}$) as $\sigma_n =\sigma_{n-1} /2$; 
\par\noindent
3) take as ``pilot function" $f_p(\vec{x})$ the density kernel estimate 
$f_k(\vec{x})$ with sizes 
$\sigma_i$ fixed at $\sigma_n$ $\forall i$ (see eq. 27 and 28 of Pisani 1993); 
\par\noindent
4) set $\sigma'_i= \lambda_i  \sigma_n$, where $\lambda_i$ is a
``local bandwidth" defined as
\begin{equation}
\lambda_i=\left( {{ f_p(\vec{x}_i)} \over {g }} \right)^{-0.5}
\end{equation}
where
\begin{equation}
\log\ g= {1 \over N} \sum_{i=1}^{N} \log(f_p(\vec{x}_i))
\end{equation}
Now, the sizes of the kernels $\sigma'_i$ are depending somehow from the 
local density through the factor $\lambda_i$: the higher is the density, the
smaller are the $\sigma'_i$; 
\par\noindent
5) evaluate the cross-validation function $CV[f_k] (\sigma_n)$ of the $n^{th}$ 
estimate of the density $f_k$ (eq. 4 and 5 of Pisani 1996).
The  $CV[f_k] (\sigma_n)$ estimates the quadratic deviation between
$f_k(\vec{x})$ and the ``true" parent density function $f(\vec{x})$ (see
eq. 33 of Pisani 1993).
\par\noindent
The steps from (2) to (5) are iterated until the minimum of $CV[f_k] (\sigma_n)$
is reached: the corresponding kernel widths $\sigma'_i$ give the
optimal density estimate $\hat{f}_{k}(\vec{x})$ as:
\begin{equation}
\hat{f}_{k}(\vec{x})=  {1\over N} \sum^N_{i=1} K(\vec{x}_i,\sigma'_i; \vec{x}) 
\end{equation}

Peaks in the density distribution function are assumed to define the presence
of clusters (in this section we use the word cluster in the statistical sense):
the local maxima of $\hat{f}_{k}(\vec{x})$ are found by solving for each 
starting $i^{th}$ galaxy position ($i=1,N$) the iterative equation 
\begin{equation}
\vec{x}_{m+1,i}=\vec{x}_{m,i} + {      d \over{ 
\sum^N_{i=1}  \left( \displaystyle{{{\nabla \hat{f}_k(\vec{x}_{m,i})}\over 
{\hat{f}_k(\vec{x}_{m,i})}}}\right)^2 
} }
\left( {{\nabla \hat{f}_k(\vec{x}_{m,i})}\over{\hat{f}_k(\vec{x}_{m,i})}}\right)
\label{eq:path}
\end{equation}
where $\vec{x}_{1,i}$ is the original position of the $i^{th}$ galaxy. 
Eq. (\ref{eq:path}) defines a path from the original position to a limiting 
point along the maximum gradient of the function $\hat{f}_k(\vec{x})$
All starting points which reach the same
limiting position $\vec{x}_{lim}$ for $m \to \infty$ 
define a cluster.
From the analysis of the distribution of the distances between 
the various $\vec{x}_{lim}$ in our samples, we decided to consider coincident 
all the limiting points that differ by less than $0.1$ arcmin.

The significance of the cluster is estimated by calculating first the likelihood
function of the sample
\begin{equation}
L= \prod_{i=1}^{N}  f_k(\vec{x}_i) = \prod_{i=1}^{N} \sum_{\mu=0}^{\nu} F_{\mu}
(\vec{x}_i)
\label{eq:likelihood}
\end{equation}
where $ F_{\mu}(\vec{x}_i)$ is the contribution to the density function of
the $\mu^{th}$ cluster $C_{\mu}$, calculated as
\begin{equation} 
 F_{\mu}(\vec{x})={1\over N} \sum_{j\in C_{\mu}} K(\vec{x}_j,\sigma_j; \vec{x})
\end{equation}
where the sum is done on the galaxies belonging to the $\mu^{th}$ cluster
and $\nu$ is the number of clusters found in the sample. For $\mu = 0$
we refer to the set of isolated galaxies, i.e. those that
did not reach a $\vec{x}_{lim}$ shared with other object.

Then the likelihood of the $\mu^{th}$ cluster ($L_{\mu}$) is calculated by 
replacing in Eq. (\ref{eq:likelihood})
the sizes $\sigma'_i$ of the kernels of the substructure objects
with $\sigma_{bck}$,  i.e. the kernel size of the background.
The ratio $\chi^2=\displaystyle{-2\ln{ L_{\mu} \over L}}$ is distributed as a 
chi square variable with one degree of freedom and is related to the 
probability that the density distribution is better described by the presence
of the $\mu^{th}$  cluster rather than by a uniform background.

In an ideal case, it would be possible to determine $\sigma_{bck}$ by using 
the isolated galaxies, because they are likely to be part of the background.
However, also local fluctuations of the density field of a cluster
can be identified as local maxima and hence this fact causes a misidentification
of cluster members as isolated galaxies.

A method to identify really isolated galaxies is to check the 
relative contribution to the total density  of the kernel of the $i^{th}$ galaxy
in the position $\vec{x}_i$ (Pisani 1996), given by
\begin{equation}
 P(i\in C_0)={1 \over N}   {  {  K(\vec{x_i},\sigma_i;\vec{x_i})} \over 
{  \hat{f}_{k}(\vec{x})  }} 
\label{eq:isolated}
\end{equation}
This equation estimates the probability that a ``nominally" isolated galaxy is 
really part of the background.
We found that for our samples $P(i\in C_0)\le 0.15$, hence we can conclude
that there are no really isolated galaxies.
Note that the estimate of  $P(i\in C_0)$ is independent from the cluster 
identification step (Eq. \ref{eq:path}). 

In order to determine a value for the background, we plotted the histogram
of the kernel sizes $\sigma'_i$, which is a symmetric distribution
with a tail at higher values. We choose as $\sigma_{bck}$ the value at the
beginning of the tail (see Fig. \ref{fig:histo}a in Sect. 3.1).

The group significances calculated from the values of $\chi^2$
are strictly correct only if the background is well determined, 
that is not our case.
For this reason we regard the $\chi^2$ only as a ranking parameter
for the existence of the groups and we consider only subclusters
in the tail at higher values of this parameter (see Fig. \ref{fig:histo}b
in Sect. 3.1).
This choice corresponds approximately to a ``formal" significance of 
$99.9\%$.
We forced the background at various reasonable densities in order
to check the robustness of our results, finding that the general clustering 
pattern of our samples is not strongly
dependent on the poorly known global background density.

As a test of the efficiency of this method, we applied DEDICA to simulated
Gaussian distributions, where no substructures are expected.
We performed various simulations
of 2-D and 3-D Gaussians, with a number of points between 300 and 700 (which
are the approximate numbers of galaxies in our various samples), finding that 
the success rate (i.e. absence of significant substructures)
of DEDICA is always between $80\%$ and $90\%$ (depending on the number of 
objects). In the cases in which DEDICA finds significant substructures, the
numbers of these ``spurious" groups is of the order of 2-3.

This method differs from the wavelet decomposition because it is non-parametric
and scale-independent. Indeed, the wavelet method is a deconvolution procedure,
which assumes an exploring function, which is fixed throughout the whole 
sample volume.
On the contrary, DEDICA adopts a kernel function, which is locally density
dependent and in this sense is not a deconvolution method (Silverman 1986).
The wavelet analysis is able to detect structures depending on the 
scale of the exploring function, while DEDICA does not requires any
assumption on the scale of clusters.
\section{The A3558 cluster complex}

In a series of papers (Bardelli et al. 1994, 1996, 1998, hereafter Paper 1, 2 
and 3; Venturi et al. 1997)
we performed a detailed multiwavelength study of the A3558 cluster complex,
a remarkable association formed by three ACO (Abell, Corwin \& Olowin 1989) 
clusters A3556, A3558 and A3562, already noticed by Shapley (1930).
This structure is approximately at the geometrical center of the Shapley 
Concentration and can be considered the core of this supercluster.
By the use of multifibre spectroscopy, we obtained 714 redshifts of galaxies 
in this region, confirming that the complex is a single connected structure,
elongated perpendicularly to the line of sight. In particular, the number of 
measured redshifts of galaxies belonging to A3558 is $307$, and thus this
is one of the best sampled galaxy clusters in the literature.
Moreover, two smaller 
groups, dubbed SC 1327-312 and SC 1329-313, were found both in optical 
(Paper 1) and X-ray band (Breen et al. 1994; Paper 2).
In particular, in Paper 2 we detected a bridge of
X-ray emitting hot gas connecting A3558 and SC 1327-313 (and possibly 
extending to A3562, see figure 7 of  Ettori et al. 1997) which resembles the 
structure between Coma and NGC4839 described by Burns et al. (1994). 
The presence of this feature strongly supports the suggestion by
Metcalfe et al. (1994) that A3562 already encountered A3558: the bridge
could be the tail of gas left by A3562 after the first passage through A3558
(for details see Burns et al. 1994).

The clusters of this complex attracted the attention of various investigators,
who applied several statistical tests in order to detect substructures.
Bird (1993), applying both 2-D and 3-D tests to galaxies in A3558, found three 
subclumps: the first coincides with the cluster center and corresponds to the 
main component, the second lies eastward of the 
cluster center and the third is in the south-east quadrant. 

In Paper 3, applying the test developed by Dressler \& Shectman
(1988), we detected a group at $\sim 10$ arcmin from the position of the second
clump of Bird (1993).
From the analysis of the isodensity contours
and the velocity distribution histograms, there was evidence of
another density excess at $\sim 15$ arcmin southward of the A3558 center.
Moreover, A3558, SC1329-313 and A3556 appear to have bimodal velocity 
distributions.

Girardi et al. (1997), using a modified version of the bi-dimensional 
wavelet decomposition which takes into account also the dynamical information, 
individuated in A3558 the same subclumps found by Bird (1993) and by us 
(Paper 3), and an additional substructure westward of the center of A3562.

Dantas et al. (1997), using the bi-dimensional wavelet decomposition
in A3558, detected a bimodality in the core of the 
cluster. The authors claim that this result resolves both the cD offset
problem (in fact the dominant galaxy was suspected not to be  at rest 
with respect to the cluster velocity centroid, Gebhardt \& Beers 1991), 
and the ``$\beta$ problem" (Paper 2), by decreasing the central
velocity dispersion from $\sim 1000$ km/s to $\sim 800$ km/s

Analysing a ROSAT PSPC image of A3558, we noted (Paper 2) 
that the diffuse emission of the hot gas is better described by two components, 
one centered on the geometrical center of the cluster and the other on the
cD galaxy. Slezak et al. (1994) applied the wavelet analysis to this
X-ray map and found evidence of a substructure nearby the cluster core.

\begin{table*}
\caption[]{Groups in the bi--dimensional sample }
\begin{flushleft}
\begin{tabular}{llllllll}
\noalign{\smallskip}
\hline\noalign{\smallskip}
  \#  & $\alpha$ (2000) & $\delta$ (2000) & \# mem.& \# vel. & $<v>$ & 
$\sigma$ &  notes \\
\noalign{\smallskip}
\hline\noalign{\smallskip}
 B109 & 13~22~31.4 & -31~18~22 & 34 & \ \ & \ \ & \ \ & 
A5556      \\
 B194 & 13~22~57.7 & -31~37~46 & 67 & 14 & 14360$^{+117}_{-145}$  & 520$^{+115}
 _{-68}$ & A3556     \\
 B353 & 13~24~15.4 & -31~40~40 & 46 & 22 & 14492$^{+131}_{-164}$  & 575$^{+86}
 _{-44}$ & A3556     \\
\multicolumn{8}{l}{                                  } \\
 B694 & 13~26~58.3 & -31~21~29 & 57 & 20 & 14038$^{+151}_{-224}$  & 771$^{+106}
 _{-138}$ & A3558     \\
 B830 & 13~27~07.2 & -31~51~53 & 72 & 26 & 14728$^{+147}_{-178}$  & 674$^{+181}
 _{-173}$ & A3558     \\
 B906 & 13~27~52.0 & -31~29~23 & 81 & 52 & 14501$^{+125}_{-183}$ & 1001$^{+119}
 _{-80}$ & A3558     \\
 B1000 & 13~28~04.0 & -31~44~41 & 55 & 24 & 13813$^{+136}_{-286}$ & 947$^{+112}
 _{-78}$ & A3558     \\
 B1014 & 13~28~05.8 & -31~05~24 & 24 & \ \ & \ \  & \ \ & A3558     \\
 B1056 & 13~28~07.4 & -31~33~05 & 53 & 29 & 14064$^{+112}_{-282}$ & 991$^{+178}
 _{-122}$ & A3558     \\
 B1199 & 13~29~00.5 & -31~49~23 & 51 & 18 & 14728$^{+259}_{-279}$  & 971$^{+122}
 _{-189}$ & A3558     \\
 B1311 & 13~29~23.7 & -31~21~34 & 57 & 23 & 15029$^{+228}_{-375}$  & 1081$^{+135}
 _{-168}$ & A3558     \\
\multicolumn{8}{l}{                                  } \\
 B1443 & 13~30~28.0 & -31~36~41 & 100 & 37 & 14652$^{+182}_{-149}$  & 842$^{+150}
 _{-56}$ & Poor     \\
 B1573 & 13~31~08.2 & -31~46~58 & 111 & 49 & 14059$^{+116}_{-257}$  & 1051$^{+160}
 _{-53}$ & Poor     \\
\multicolumn{8}{l}{                                  } \\
 B1952 & 13~33~28.1 & -31~41~16 & 179 & 35 & 14447$^{+230}_{-305}$  & 1113
$^{+103}_{-112}$ & A3562     \\
 B2103 & 13~34~50.8 & -31~40~14 &37 &  8  & 14751$^{+156}_{-502}$ & 880$^{+396}_{-107}$ & 
A3562     \\
 B2240 & 13~35~38.3 & -31~55~40 & 76 & 18 & 14343$^{+445}_{-100}$ & 1462$^{+445}_{-100}$ & 
A3562   \\
\noalign{\smallskip}
\hline
\end{tabular}
\end{flushleft}
\label{tab:bi-dim}
\end{table*}

\subsection{The bi--dimensional sample}

\begin{figure}
\epsfysize=5cm
\epsfbox{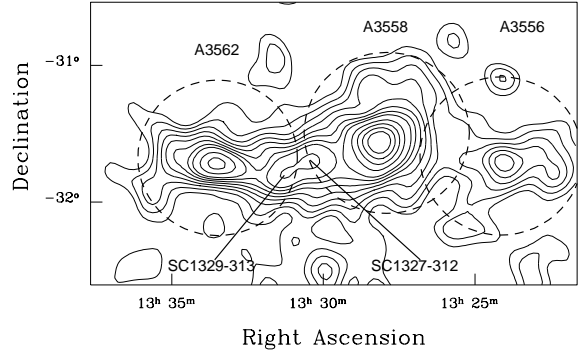}
\caption[]{Galaxy isodensity contours of the region of the A3558 cluster 
complex: the objects are binned in $2$ arcmin $\times$ $2$ arcmin 
pixels and then smoothed with a Gaussian with a FWHM of $6$ arcmin. Dashed
lines correspond to 1 Abell radius circles around cluster centers. The positions
of the two poor clusters SC 1327-312 and SC 1329-313 are shown.}
\label{fig:iso}
\end{figure}

From the COSMOS/UKST galaxy catalogue (Yentis et al. 1992), we extracted a 
rectangular region of $3^o.2\times 1^o.4$, corresponding to 
$13^h 22^m 06^s<\alpha(2000) < 13^h 37^m 15^s$ and 
$-32^o 22' 40''<\delta(2000) < -30^o 59' 30''$. 
This region is part of the UKSTJ plate $444$ and contains the A3558 complex.
We restricted our analysis to the $2241$ galaxies with magnitudes brighter
than $b_J=19.5$. The isodensity contours of galaxies of this region are
presented in Fig. \ref{fig:iso}.
\par

\begin{figure}
\epsfysize=8.5cm
\epsfbox{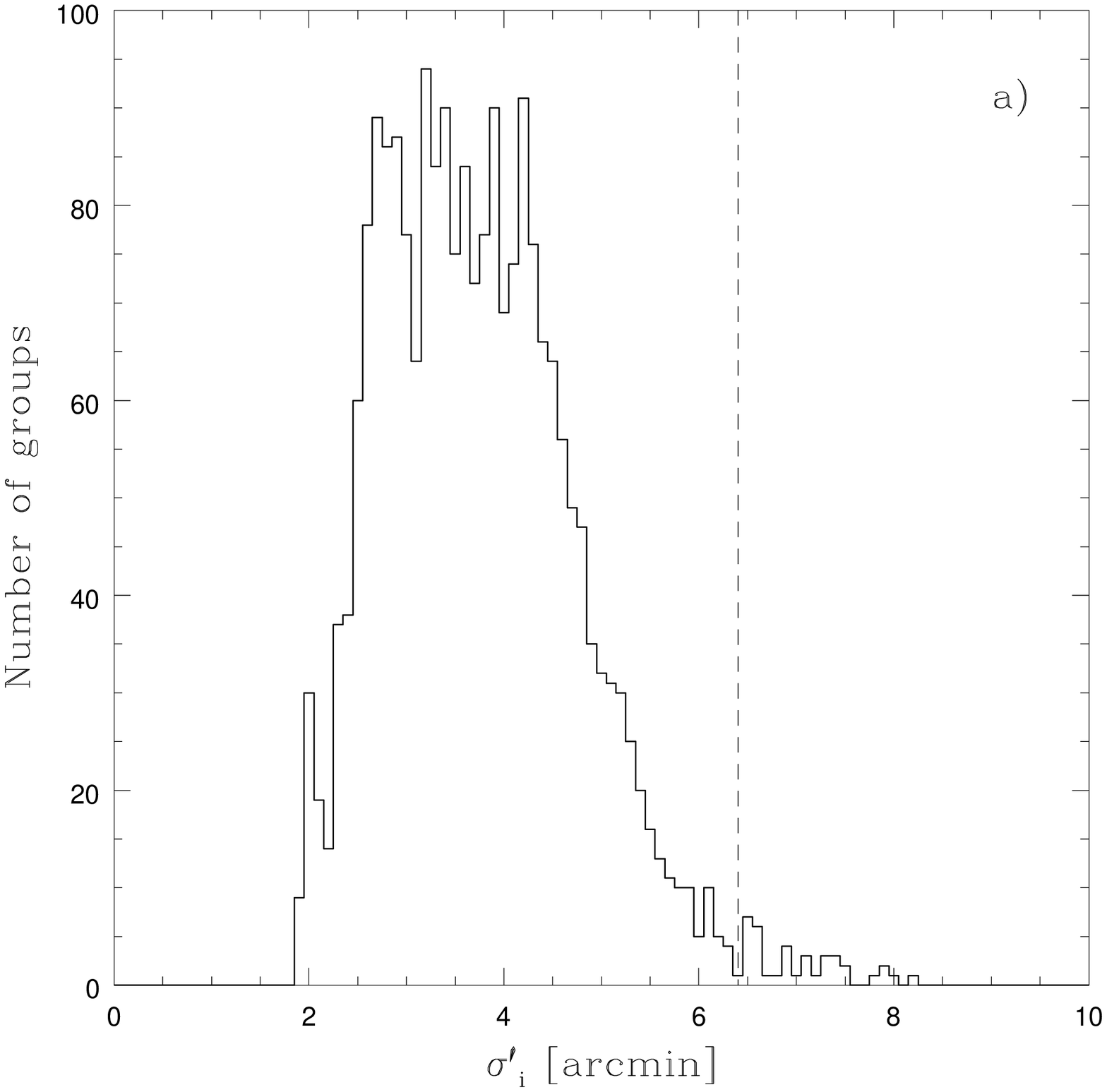}
\epsfysize=8.5cm
\epsfbox{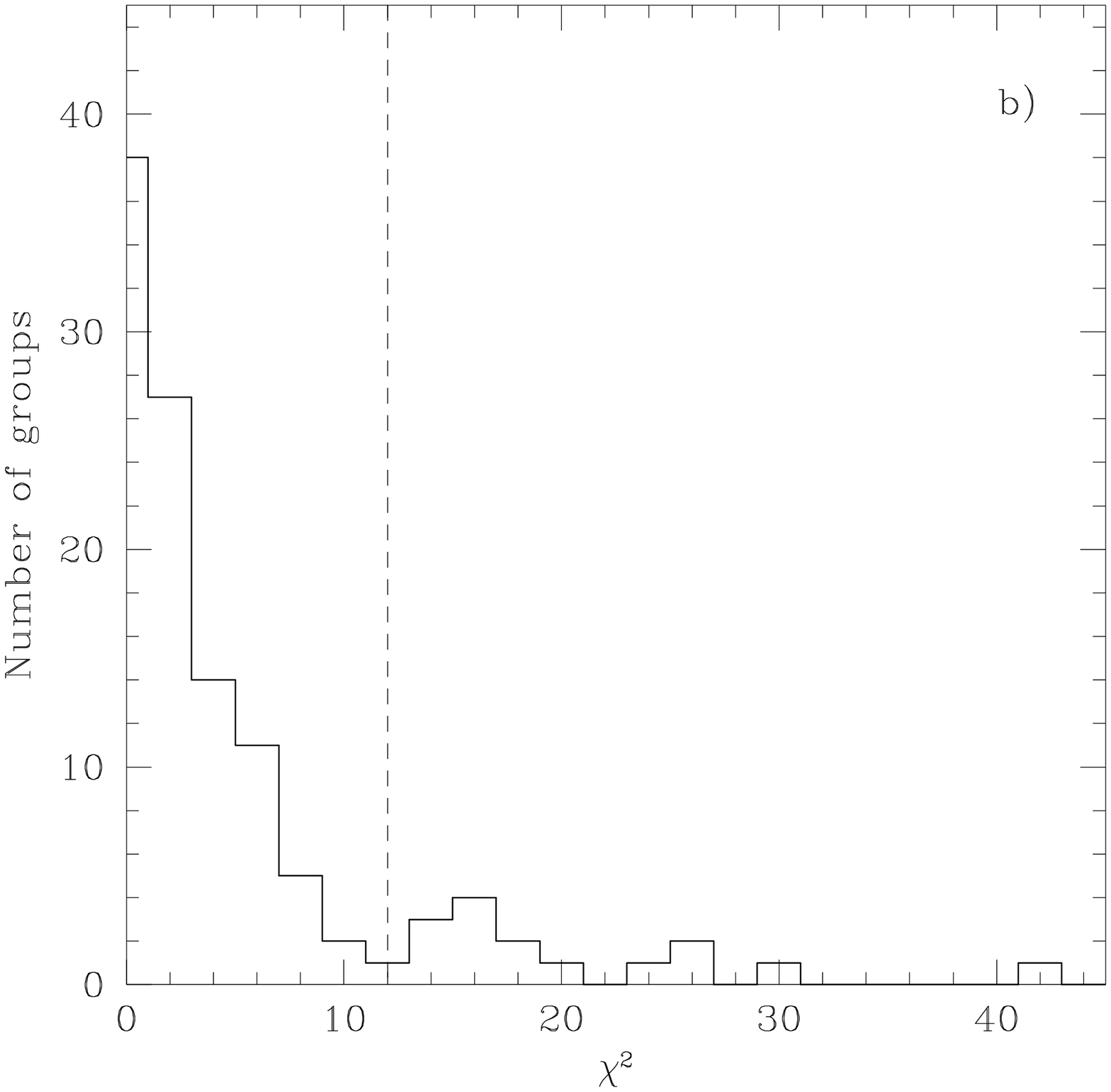}
\caption[]{ a) Histogram of the kernel sizes $\sigma'_i$ for the bi--dimensional
sample. The vertical dashed line corresponds to the adopted $\sigma_{bck}=6.4$
arcmin. b) Histogram of the $\chi^2$ values for the detected groups in the
bi--dimensional sample. The vertical dashed line corresponds to the adopted 
threshold of significance ($\chi^2 = 12$). }
\label{fig:histo}
\end{figure}

The adopted $\sigma_{bck}$ is 6.4 arcmin, which corresponds to 
$0.05$ gal arcmin$^{-2}$, and a $\chi^{2}$ threshold of 12
(Fig. \ref{fig:histo}). 

The DEDICA algorithm found 113 groups, 16 of which are considered
significant.
In Table
\ref{tab:bi-dim} the 16 significant clusters sorted by right ascension are 
listed. 
Column (1) is the identification number while columns (2) and (3) give the 
$\alpha$ and $\delta$
coordinates of the group center. We identified these points 
as the common $\vec{x}_{lim}$ position of the members and they do not
necessarily coincide with the geometrical centers. Columns (4) and (5) 
report the number of the cluster members
and of the available velocities in the group. Columns (6) and (7) 
are the dynamical parameters computed using the biweight estimators
of position and scale (see Paper 1 for details on the method).
Finally, Column (8) reports the main component to which
the substructures belong, i.e.
the A3556, A3562, A3558 Abell clusters, and the poor clusters SC 1317-312
and SC1329-313 (denoted as ``Poor").

The velocity sample (see below) covers an area which is $\sim 17 \%$ of the
bi-dimensional sample and therefore not all subclusters have measured 
velocities: for this reason in two cases the dynamical parameters are 
left blank in Table \ref{tab:bi-dim}.

\subsection{The three--dimensional sample }

In this case we used the velocity sample described in Paper 3. 
From the $714$ galaxies with measured redshift we selected the $600$ objects in
the velocity range $[10000-20000]$ km/s. The velocity sampling of the
A3558 complex is not constant: it varies from $\sim 20 \%$ in the region
of A3562 to $\sim 48\%$ in the core of A3558. 
This fact could cause a bias that erases undersampled substructures or
leads to false detections in regions more sampled than the average.

In order to check this point, we performed the bi-dimensional 
analysis on this subsample. Among the 14 bi-dimensional groups
present in the area of the velocity sample, 13 are correctly detected
and $B2240$ (corresponding to T599)
has been lost because of a significance lower than the limit.
The tight correspondence between bi-dimensional and  these
``three-dimensional" groups indicates that this bias is not severe. 

Given the fact that the velocities are not related to positions because of
the dominance of peculiar motions over the Hubble flow,
a problematic point is how to handle the velocity coordinate. Our formalism
uses symmetric three-dimensional gaussian kernels, with $\sigma_x=\sigma_y=
\sigma_v$ and therefore it is necessary to scale the velocity interval
in order to have a numerical range comparable to the other two variables.
The typical $\sigma$ of the spatial coordinates is 2-3 arcmin, while
the typical $\sigma_v$ (determined on the one-dimensional redshift 
distribution) is $\sim 200$ km/s.
We have therefore chosen to compress the velocity by a factor 100;
we have also checked that the method gives essentially the same results
if this reduction factor is in the range $50-500$.

A more precise approach to this problem requires the modelling of the peculiar 
velocity field. This implies the detailed knowledge of the mass distribution
in the A3558 complex, which is presently not available.
In fact there is no general agreement even for the mass of the dominant,
best studied, cluster of the complex (A3558). 
The mass obtained from optical data ranges from $3.4 \times 10^{14}$  
h$^{-1}$ M$_{\odot}$ (Dantas et al. 1997) to $6 \times 10^{14}$  
h$^{-1}$ M$_{\odot}$ (Biviano et al. 1993), while the X-ray data
give mass in the range $1.5-6.0 \times 10^{14}$  
h$^{-1}$ M$_{\odot}$ (Paper 2, Ettori et al. 1997).

The adopted value of the background is $\sigma_{bck}=7.5$ arcmin, corresponding 
to 0.003 gal arcmin$^{-2}$ (km/s/100)$^{-1}$ and a significance threshold
of $\chi^2=20$.
The significant 3-D clusters are $21$, 
out of a total of $36$, and are reported in Table \ref{tab:three-dim}. 
Column (1) is the subcluster identification number, Columns (2), (3) and (4) 
report the coordinates ($\alpha$, $\delta$ and $v_o$) of the group center, 
while the number of members is reported in Column (5).
The estimated dynamical parameters are reported in Columns (6) and (7) and
in Column (8) the general position is given. 
Note that in some cases, the velocity distributions are highly asymmetric and
therefore there is a significant difference between $v_o$ (the position
of the density peak) and the mean velocity estimated with the biweight method. 

\begin{figure}
\epsfysize=8.5cm
\epsfbox{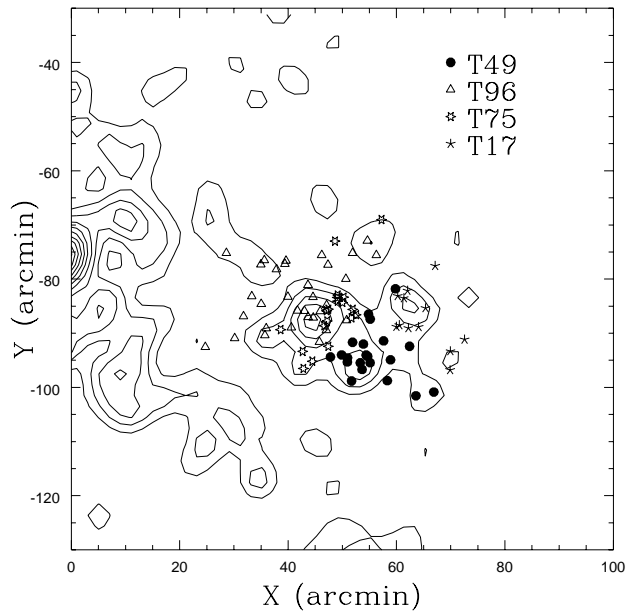}
\caption[]{Groups in the A3556 cluster region. The positions of sub-cluster
galaxies found in the three-dimensional sample are overplotted to the 
2-D isodensity contours smoothed with a 2 arcmin FWHM Gaussian. 
Symbols used to label the different groups are explained in the figure.}
\label{fig:a3556}
\end{figure}

\begin{figure}
\epsfysize=8.5cm
\epsfbox{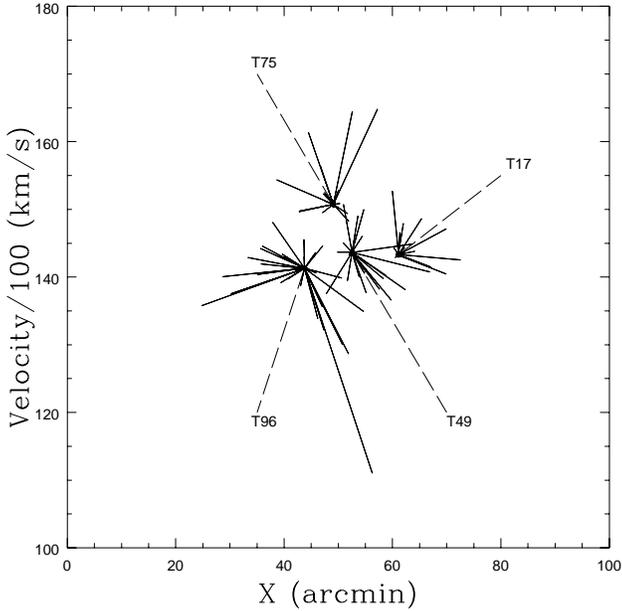}
\caption[]{Projection on the X-Velocity plane of groups in the A3556 cluster
region. Solid lines connect the position of each galaxy to the common limiting 
position $\vec{x}_{lim}$ of the group.}
\label{fig:a56xv}
\end{figure}

\begin{figure}
\epsfysize=8.5cm
\epsfbox{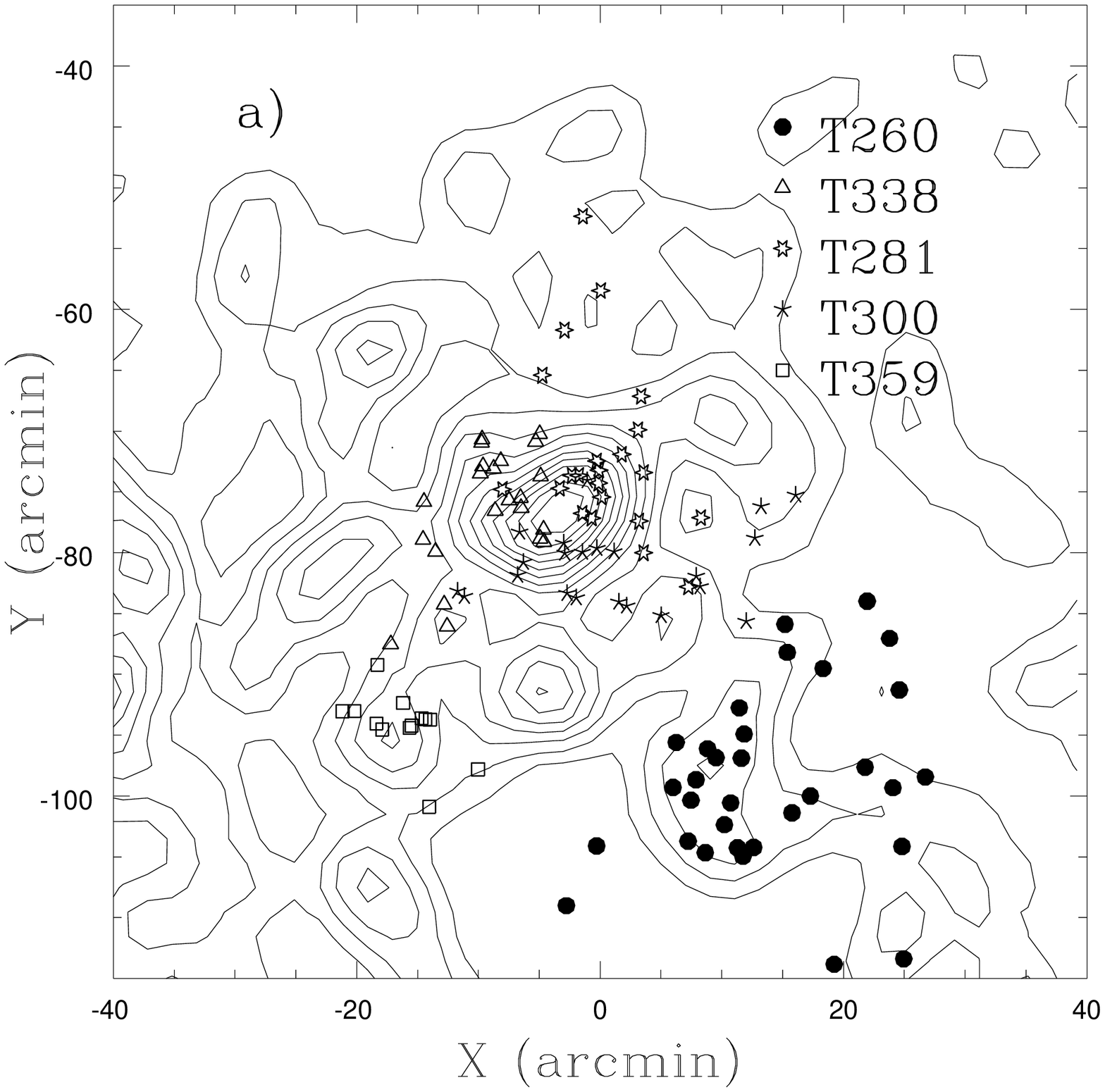}
\epsfysize=8.5cm
\epsfbox{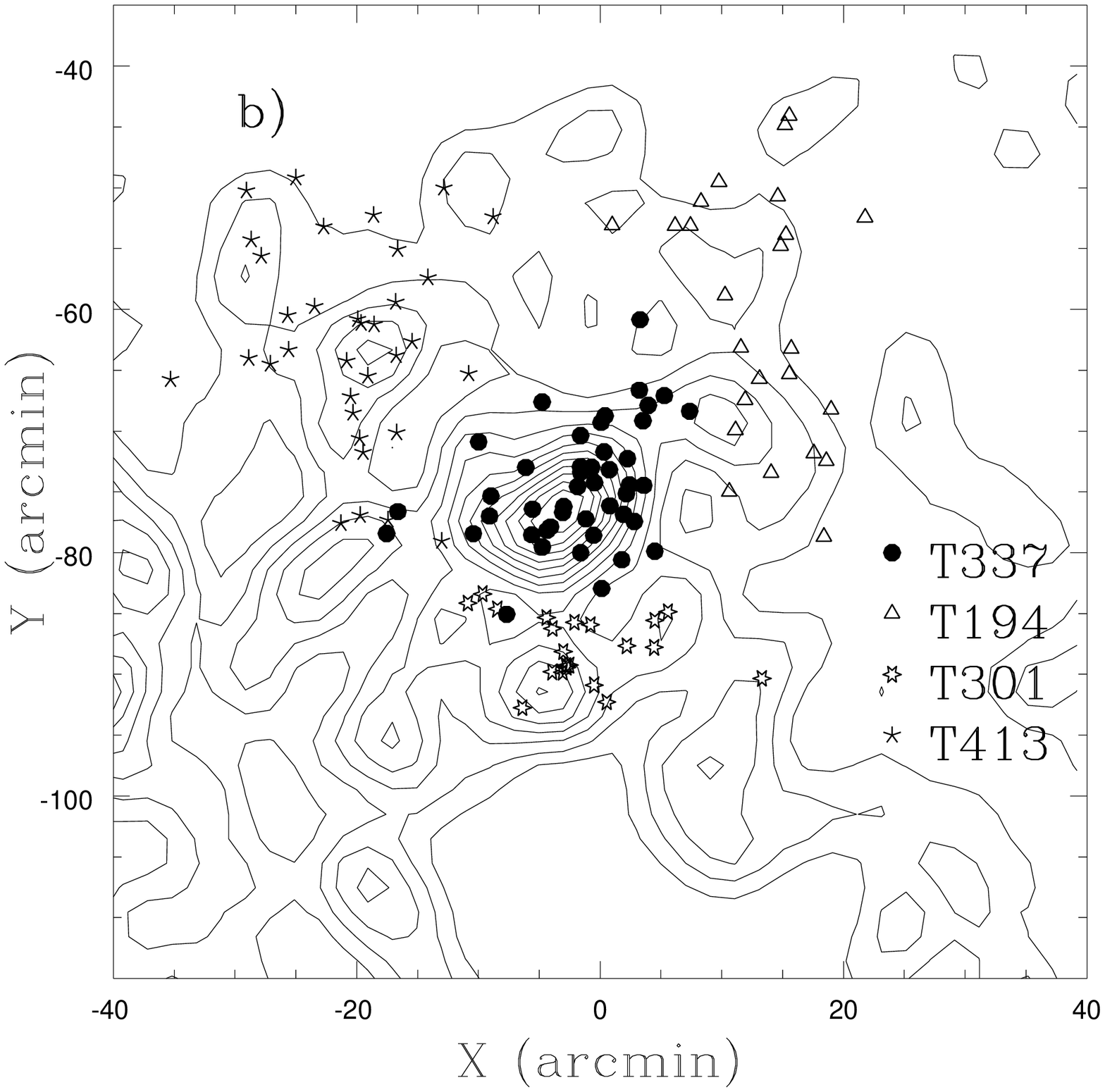}
\caption[]{The same as Fig. \ref{fig:a3556}, for groups in the A3558 cluster 
region. For clearness the group members have been split in two panels}
\label{fig:a3558}
\end{figure}

\begin{figure}
\epsfysize=8.5cm
\epsfbox{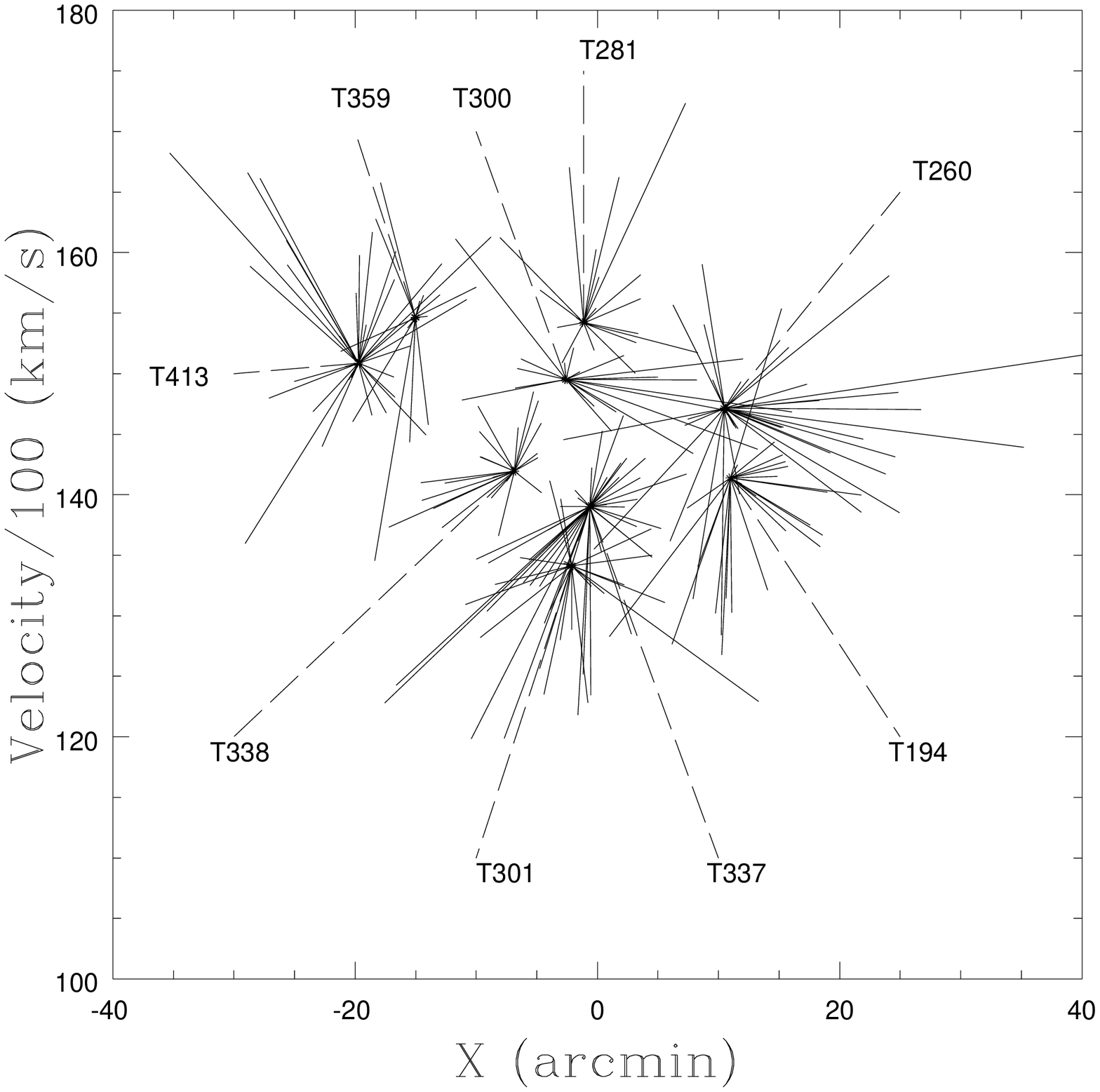}
\caption[]{The same as Fig. \ref{fig:a56xv} for groups in the A3558 cluster
region. }
\label{fig:a58xv}
\end{figure}

\begin{figure}
\epsfysize=8.5cm
\epsfbox{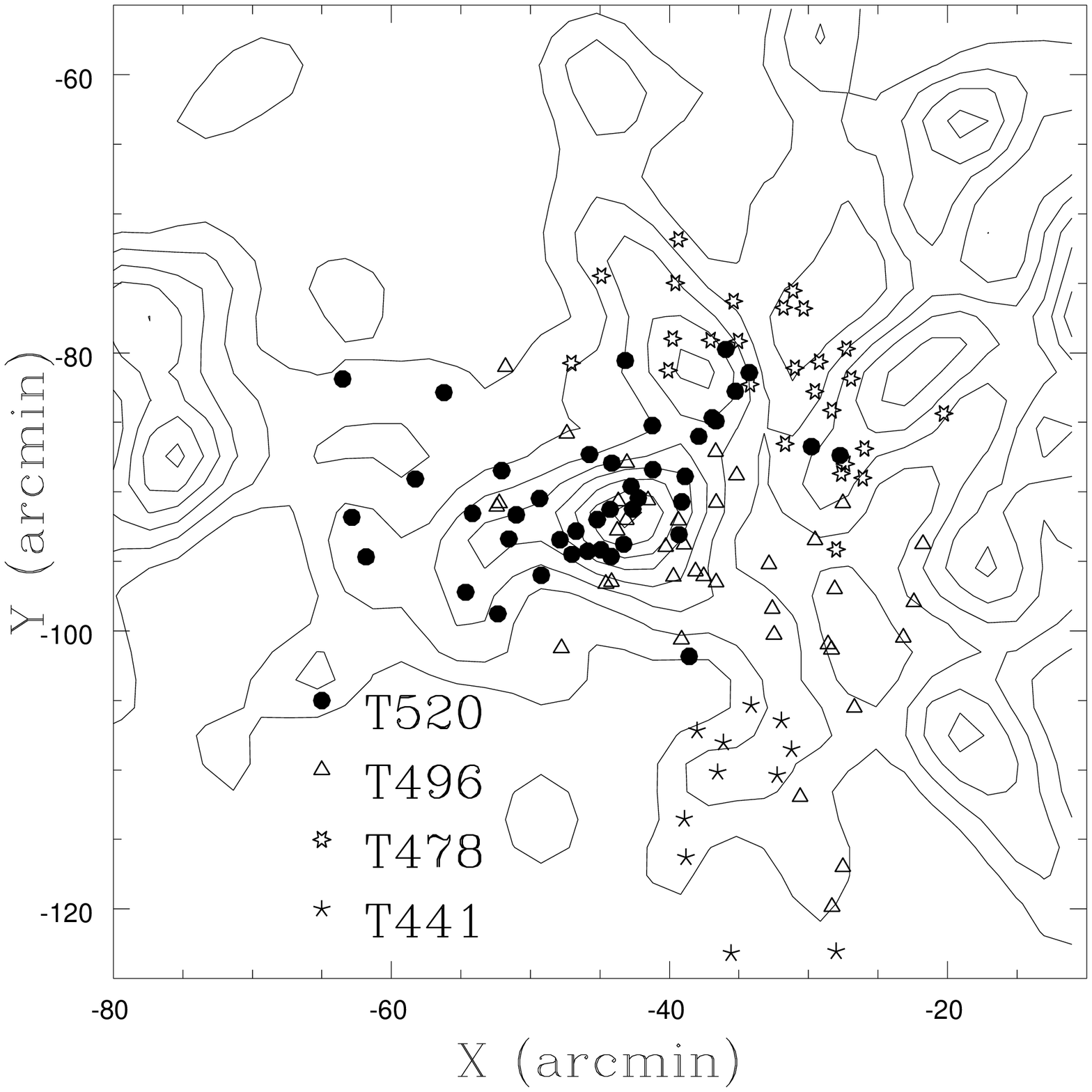}
\caption[]{The same as Fig. \ref{fig:a3556}, for groups in the Poor clusters
region. }
\label{fig:poor}
\end{figure}

\begin{figure}
\epsfysize=8.5cm
\epsfbox{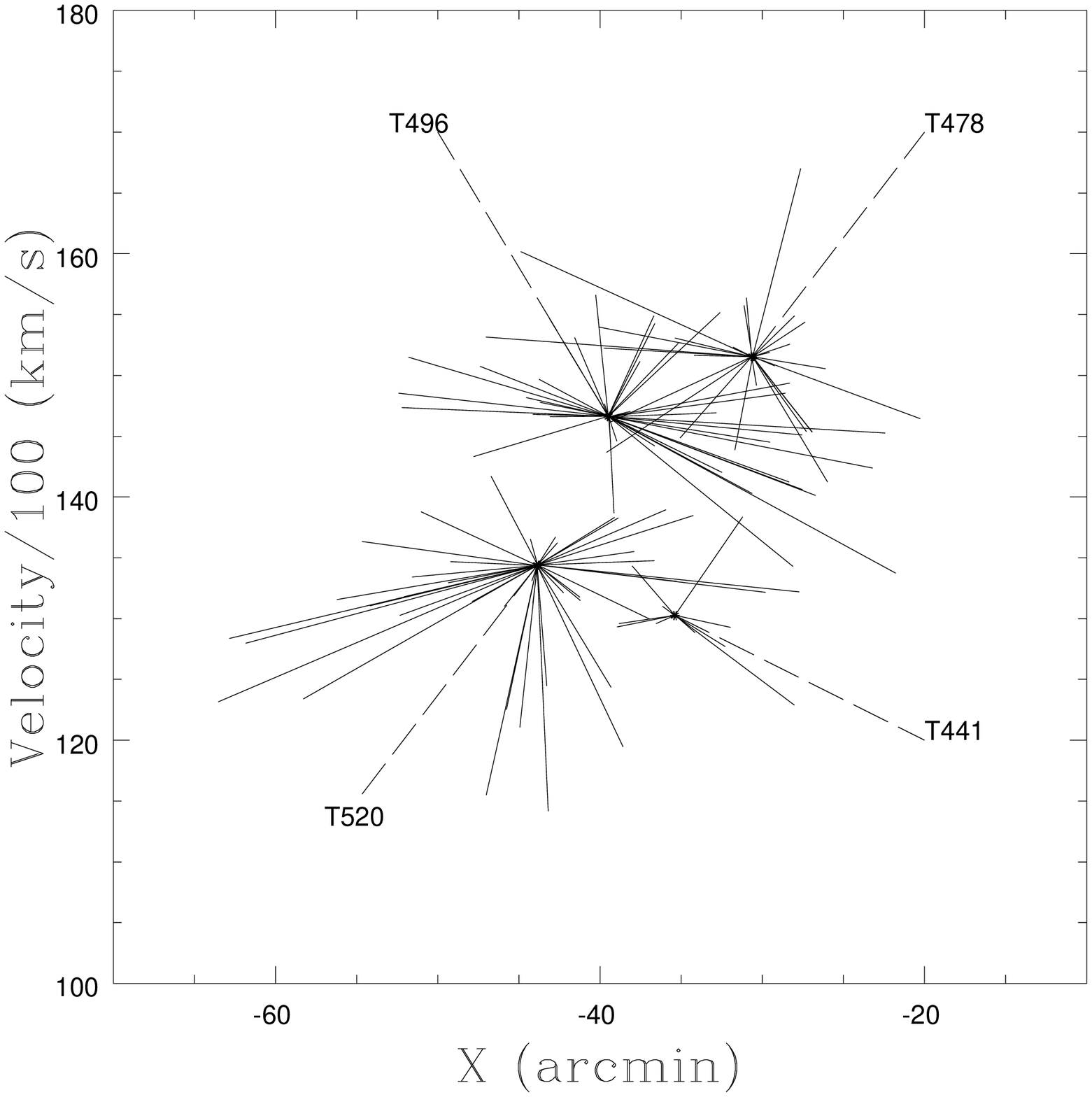}
\caption[]{The same as Fig. \ref{fig:a56xv} for groups in the Poor cluster
region. }
\label{fig:poxv}
\end{figure}

\begin{figure}
\epsfysize=8.5cm
\epsfbox{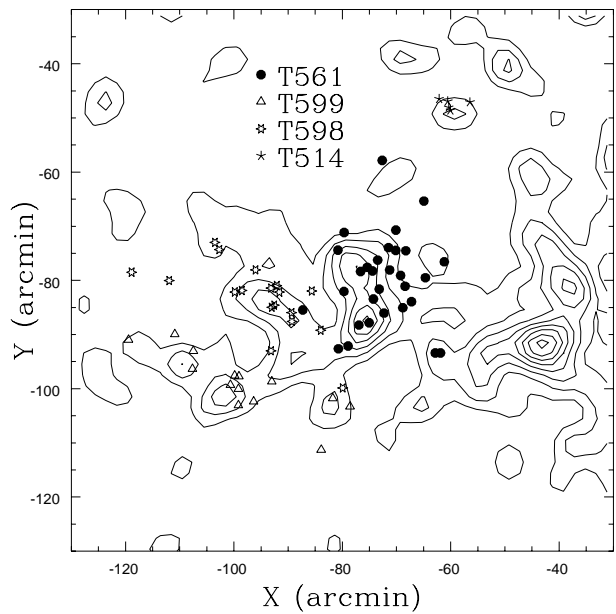}
\caption[]{The same as Fig. \ref{fig:a3556}, for groups in the A3562 cluster
region.}
\label{fig:a3562}
\end{figure}

\begin{figure}
\epsfysize=8.5cm
\epsfbox{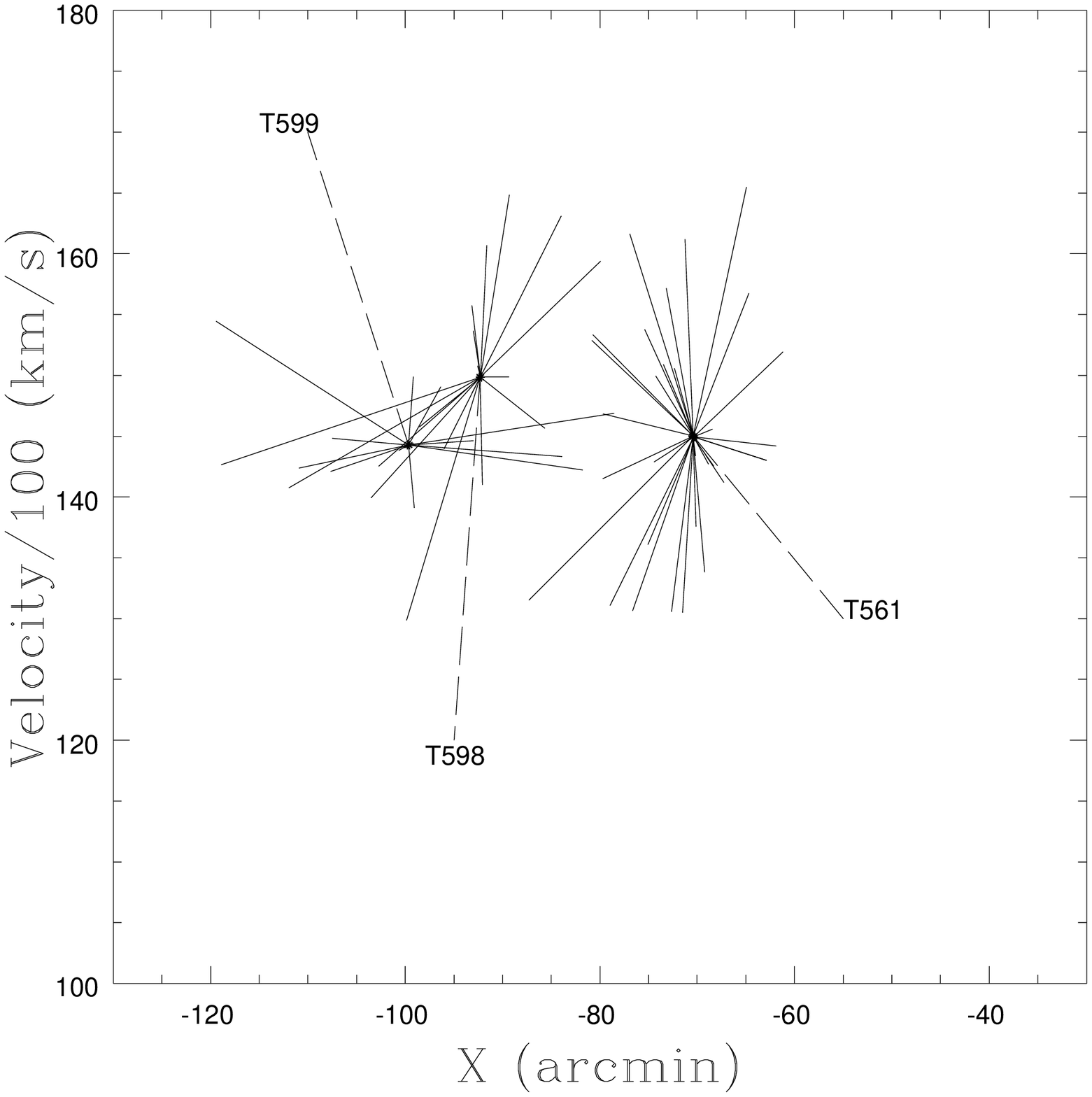}
\caption[]{The same as Fig. \ref{fig:a56xv} for groups in the A3562 cluster
region. }
\label{fig:a62xv}
\end{figure}

In Figs. \ref{fig:a3556}, \ref{fig:a3558}, \ref{fig:poor} 
and \ref{fig:a3562} the positions of the substructure members are shown,
superimposed on the isodensity contours of galaxies obtained with a smoothing
of $2$ arcmin. Note that in these figures the isodensity contours 
are obtained directly by binning the data in one arcmin cells
and not from the $\hat{f}_k$: however the two methods give similar
contours.
In Figs. \ref{fig:a56xv}, \ref{fig:a58xv}, \ref{fig:poxv} and 
\ref{fig:a62xv}, the groups members are shown projected on the X-Velocity 
plane. Solid lines connect the position of each galaxy to the common limiting 
position $\vec{x}_{lim}$ of the group.

\begin{table*}
\caption[]{ Three--dimensional sample }
\begin{flushleft}
\begin{tabular}{llllllll}
\noalign{\smallskip}
\hline\noalign{\smallskip}
  \#  & $\alpha$ (2000)  & $\delta$ (2000) & $v_0$ &\# mem. & $<v>$ & $\sigma$ 
  & notes \\
\noalign{\smallskip}
\hline\noalign{\smallskip}
 T17 & 13~23~03.0 & -31~43~37 & 14501 & 12 & 14498$^{+83}_{-117}$ & 
334$^{+124}_{-55}$  & A3556     \\
 T49 & 13~23~42.7 & -31~49~23 & 14366 & 21 & 14173$^{+104}_{-125}$ & 
410$^{+68}_{-69}$  & A3556       \\
 T75 & 13~23~59.0 & -31~39~57  & 15075 & 19 & 15087$^{+28}_{-71}$ & 
154$^{+34}_{-34}$ & A3556     \\
 T96 & 13~24~23.9 & -31~41~14  & 14129 & 32 & 14035$^{+58}_{-125}$ & 
397$^{+61}_{-46}$ & A3556     \\
\multicolumn{8}{l}{                                  } \\
 T194 & 13~26~56.9 & -31~22~34 & 14139 & 24 & 13721$^{+172}_{-111}$ & 
625$^{+178}_{-52}$  & A3558     \\
 T260 & 13~26~59.0 & -31~54~46 & 14715 & 34 & 14605$^{+109}_{-88}$ & 
622$^{+134}_{-133}$  & A3558     \\
 T337 & 13~27~50.9 & -31~28~48 & 13903 & 45 & 13597$^{+90}_{-210}$ & 
693$^{+163}_{-40}$  & A3558     \\
 T281 & 13~27~53.3 & -31~30~47 & 15420 & 22 & 15456$^{+83}_{-56}$ & 
294$^{+68}_{-39}$  & A3558     \\
 T301 & 13~27~57.9 & -31~44~35 & 13416 & 21 & 13224$^{+113}_{-152}$ & 
500$^{+75}_{-67}$  & A3558     \\
 T300 & 13~28~00.1 & -31~35~08 & 14952 & 22 & 14876$^{+53}_{-82}$ & 
242$^{+46}_{-36}$  & A3558     \\
 T338 & 13~28~19.8 & -31~31~31  & 14195 & 22 & 14213$^{+63}_{-93}$ & 
345$^{+56}_{-14}$  & A3558     \\
 T359 & 13~28~58.2 & -31~48~57 & 15463 & 13 & 15393$^{+175}_{-187}$ & 
788$^{+211}_{-202}$  & A3558      \\
 T413 & 13~29~19.9 & -31~23~16 & 15088 & 34 & 15392$^{+121}_{-151}$ & 
691$^{+110}_{-41}$  & A3558     \\
\multicolumn{8}{l}{                                  } \\
 T478 & 13~30~10.8 & -31~36~29 & 15154 & 26 & 15069$^{+80}_{-162}$ & 
515$^{+125}_{-70}$  & Poor     \\
 T441 & 13~30~33.4 & -32~03~59 & 13025 & 11 & 12948$^{+41}_{-52}$ & 
159$^{+276}_{-89}$  & Poor     \\
 T496 & 13~30~52.3 & -31~50~01 & 14663 & 38 & 14690$^{+101}_{-99}$ & 
537$^{+87}_{-32}$  & Poor     \\
 T520 & 13~31~12.8 & -31~46~29 & 13437 & 42 & 13280$^{+80}_{-95}$ & 
482$^{+87}_{-49}$  & Poor     \\
 T514 & 13~32~26.7 & -31~02~58 & 13543 & 5 & 13510$^{+248}_{-152}$ & 
287$^{+83}_{-24}$  & Poor     \\
\multicolumn{8}{l}{                                  } \\
 T561 & 13~33~31.6 & -31~35~19 & 14501 & 30 & 14527$^{+168}_{-229}$ & 
987$^{+116}_{-67}$  & A3562     \\
 T598 & 13~34~58.6 & -31~38~59 & 14981 & 18 & 14867$^{+318}_{-231}$ & 
919$^{+149}_{-141}$  & A3562     \\
 T599 & 13~35~33.6 & -31~54~24 & 14426 & 14 & 14412$^{+111}_{-50}$ & 
278$^{+106}_{-42}$  & A3562     \\
\noalign{\smallskip}
\hline
\end{tabular}
\end{flushleft}
\label{tab:three-dim}
\end{table*}
\section{Results and discussion}

\subsection{A3556}

The bi-dimensional analysis found $2$ significant groups in the inner part of 
this cluster, being the substructure $B 109$ at about one 
Abell radius from the center. The distribution of galaxies is not smooth
as can be seen also in Fig. \ref{fig:a3556}, where the isodensity contours
show three peaks. However, the dynamical analysis of the substructures reveals 
that mean velocities and  velocity dispersions are consistent with the values 
of $<v>=14357^{+76}_{-76}$ km/s and $\sigma=643^{+53}_{-43}$ km/s
given in Paper 3 for the whole cluster. 

The three-dimensional analysis shows four subclumps: two of them
($T 49$ and $T 96$) have dynamical parameters consistent each other at 
about one sigma and correspond to two 
nearby ($\sim 12$ arcmin) density maxima separated on the plane of the sky. 

The group $T 75$ appears superimposed along the line of sight
with $T 96$ (they correspond to $B353$), but at higher velocity: the presence 
of these two
peaks in the redshift distribution was already detected in Paper 3, with
consistent dynamical parameters. Moreover, as noted by Venturi et al. (1997, 
1998), $T 75$ contains the head-tail radiosource $J1324 -3138$: the head-tail
appearance of this source could indicate an interaction with a gaseous medium 
during the infall of the group toward the main clump of A3556.

The group $T 17$ seems to be at an intermediate mean velocity and separated 
in the plane of sky from the other clumps. The westernmost galaxy of this group
(well outside the density peak) corresponds to the Wide Angle Tail
radio source studied by Venturi et al. (1997). In the bi-dimensional analysis 
this object is associated to $B109$, but also in this case it is at the edge 
of the substructure. This fact is anomalous for Wide Angle Tail sources, found
normally at the center of clusters or groups (Owen 1996).

In conclusion A3556, which appears at a first look as a relatively relaxed 
cluster (see figures 1 and 2 of Paper 3), revealed after a more detailed 
analysis a relevant clumpiness in its core. This is an indication that the 
dynamical processes have not had enough time to destroy the substructures and 
to smooth the galaxy distribution.

\subsection{A3558}

The bi-dimensional analysis revealed that the inner core ($r<0.75$ \hmpc)
of this cluster could be separated in two groups: $B 906$, which contains the
dominant galaxy, and $B 1056$. This spatial bimodality was already found by 
Dantas et al. (1997) using the bi-dimensional wavelet analysis: 
the positions of their 
clumps A and B differ only by $\sim 1$ arcmin from our groups and there
is a good consistency in the dynamical parameters. However, the 
three-dimensional analysis shows that the situation is more complex, with four 
subclusters
($T337$, $T281$, $T300$ and $T338$) almost aligned along the line of sight.
In particular, the bi-dimensional group $B906$ is broken into two components,
at $<v>=15456$ km/s ($T281$) and $<v>=13597$ km/s ($T337$): note that 
the mean velocity of $T281$ is consistent with that of clump A$'$ of
Dantas et al. (1997).

The main three-dimensional structure is the group $T337$. The group 
$T194$ has similar dynamical parameters and could therefore be considered 
an elongation of $T337$. 
Merging these two clumps together in order to have a larger sample,
 we find $<v>=13640^{+78}_{-124}$ km/s 
and $\sigma=665^{+86}_{-69}$ km/s. We are consistent with  the results found 
by Bird (1994), Girardi et al. (1997) and Dantas et al. (1997) that the 
velocity dispersion of the main component of $A3558$ is of the order of 
$\sim 700$ km/s. 
This velocity dispersion is more consistent with the lower X-ray temperature 
found by ROSAT ($3.25$ keV Paper 2), rather than the higher temperature 
detected by ASCA ($5.5$ keV, Markevitch \& Vikhlinin 1997).
The velocity histogram of $T337 + T194$ is asymmetric with a tail toward low 
velocities: the peak of the distribution is located at $\sim 14100$ km/s, a 
value consistent with the cD velocity.

The bi-dimensional clump $B 1000$ has a positional coincidence in the plane of 
the sky with $T301$. These structures are located on a density excess,
southward of the cluster center, already noted in Paper 3 and detected by the 
use of the West et al. (1988) symmetry test. 
This density excess was revealed also by the presence of two peaks in
the velocity distribution in the third radial shell of Paper 3. 
The dynamical parameters of the lower velocity peak are well consistent with 
those of $T 301$, while the higher velocity peak is probably associated to 
galaxies of group $T 413$ which also fall in this shell.
The latter clump has been detected by Bird (1994), Girardi et al. (1997) and 
in Paper 3 using the Dressler \& Shectman (1988) 
$\Delta$ test. As noted in Paper 3, this substructure is in the region
of A3558 which show a significant enhancement of temperature (Markevitch \&
Vikhlinin 1997). 

Another relevant condensation is the group $T 260$ (or $B 830$), in the 
south-west part of A3558, whose isodensity contours appear elongated.

\subsection{SC1327-312 and SC1329 -313}

 An interesting feature of the A3558 complex is the presence of a number
of clumps between A3558 and A3562, which form a series connecting
the two clusters, starting with $T 413$ and $T 359$ just on the east of 
A3558. 
In the X-ray band there are two relevant diffuse emissions dubbed
SC 1327-312 and SC 1329-314 (Breen et al. 1994; Paper 2) 
already noted also in the optical band (Paper 1). 
The SC 1327-312 group probably corresponds to $T478$, 
even if there is an offset of
$\sim 5$ arcmin between the two positions. 
In Paper 3 we noted a bimodality in the velocity distribution of SC 1329-313:
these two groups are recovered as $T 520$ and $T 496$. 


\subsection{A3562}

From the analysis of the three-dimensional sample it is possible to find
out three components. The main cluster component appears to be $T561$, 
with a velocity
dispersion of $\sim 900$ km/s, consistent with that found in Paper 3. Note the
presence of other two clumps ($T599$ and $T598$) probably connected with the 
main cluster.

\section{Conclusions and Discussion}

The DEDICA algorithm is a non-parametric, scale independent method for
substructure analysis. We used this algorithm on both a bi-dimensional and
a three-dimensional (i.e. using the galaxy velocity as third coordinate) sample
in the region of the A3558 complex. The bi-dimensional analysis
finds a large number of clumps in correspondence of evident
density peaks, but it appears insufficient to 
correctly characterize all the substructures: in 
fact the three-dimensional analysis finds several cases in which 
different subclumps aligned along the 
line of sight are seen as single bi-dimensional peaks. 

 On the other hand, the not uniform and incomplete sampling in velocity 
may in principle lead to miss some structures in the three-dimensional
analysis. However, this incompleteness does not seem 
to introduce significant artifacts, because there is a good correspondence 
between the three-dimensional 
subcluster centers and most of the bi-dimensional density excesses. 

A number of these subclusters were already detected in Paper 3 by the 
simultaneous use of various techniques, such as the classical shape estimators,
the West et al. (1988) and the Dressler \& Shectman (1988) tests and the
analysis of the velocity histograms with the KMM algorithm (Ashman,
Bird \& Zepf 1994). The efficiency of these methods depends on the
properties of the substructures: for example, the West et al. test recovers
groups as departures from the mirror symmetry of the main cluster, while
the Dressler \& Shectman method is insensitive to subclusters which 
are superimposed along the line of sight or with similar mean velocity
and velocity dispersion. This fact led Pinkney et al. (1996) to recommend
for the substructure analysis of clusters the use of a wide battery
of statistical tests. Under this aspect, the DEDICA algorithm is more efficient 
and faster and permits a direct estimate of the
significance of the detected clump, without relying on long computer 
simulations.

A direct comparison with the results obtained with the wavelet decomposition
analysis is not straightforward because of the different starting samples.
The subclusters detected by Girardi et al. (1997) are a subset of our sample, 
revealing that the two methods are
similar as discussed by Fadda et al. (1998). The difference in the dynamical 
parameters are probably due to the fact that the method of Girardi et al. is not
fully three-dimensional.
Our bi-dimensional analysis of the center of A3558 gives the same
results of Dantas et al. (1997), i.e. the inner 0.5 Abell radius region
is formed by two clumps, one of which ($B 906$) approximatively centered 
(within $\sim 1$ arcmin) on the X-ray excess of the cD galaxy.
On the contrary, our three-dimensional analysis reveals a more complicated 
situation.


It is possible to consider two main hypotheses
for the origin of the A3558 cluster complex: one is that we are observing a 
cluster-cluster collision, just after the first core-core encounter, while 
the other considers repeated, incoherent group-group and group-cluster mergings
focused  at the geometrical center of a large
scale density excess (the Shapley Concentration).

In the first scenario, we could speculate that a cluster collided onto A3558
and its remnants are visible as the overdensity regions of SC 1327-312, 
SC1329-313 and A3562: the connection of hot gas between A3558 and
SC1327-312 found in Paper 2 could be a trace of such a collision. 
In this framework, A3556 would be formed by the members of the intervening 
cluster remained in the backside part of the merging direction.   
Therefore,  galaxies outside A3558 would belong to the destroyed cluster
and to the outer regions of the more massive target cluster
and form the clumpiness expected in the simulations.

Caldwell \& Rose (1997) suggested that the infall of a group onto a cluster
may trigger starburst phenomena on its galaxies. Indeed they found 
a number of objects showing recent star formation in the filament
connecting the Coma cluster with NGC4839, which was suspected by
 Burns et al. (1994) to have already passed the core of Coma.
In order to check if the cluster interaction modifies the galaxy spectral 
emission, we cross-correlated our velocity sample with the photometric
sample of Metcalfe et al. (1994), which contains $B$ magnitudes and 
the $(U-B)$, $(B-R)$ and $(R-I)$
colours to $B=20.8$. The use of the velocity sample, limited in the
$[10000-20000]$ km/s range, eliminates the ambiguity 
of possible colour effects due to objects at various redshifts. 
We consider ``blue galaxies" those with $(U-B) <0.3$: this choice
should allow to exclude the ``red sequence" in
the colour-magnitude planes typical of the presence of clusters and to have
at the same time
a large  number of objects (157) in the blue sample. 

Dividing the galaxies  into two groups, i.e. those that are in
 significant groups and those that are outside them, we found that 
the fractions of blue objects are $25 \%$ and $45 \%$, respectively.
The reddest group is $T337$ with a fraction of $5\%$ of blue
galaxies and this supports the idea 
that the group could be considered as a preexisting, already relaxed cluster
(that is A3558).
The bluest groups are $T520$  and $T441$ with $47 \%$ and $45 \%$
of blue galaxies and are in the
expected position in the scenario of a cluster-cluster collision, 
i.e. between A3562 and A3558.

In the second scenario, instead of a single merging phenomenon, there would
have been a series of minor mergings and collisions, confined in a relatively 
small region
by the deep potential well of the supercluster. In this case the different
``colours" of the groups would be related only to the different initial
galaxy  morphological composition of the merging components.

The determination of the density excess in galaxies in the entire supercluster
(Bardelli et al., in preparation), through the analysis of a inter-cluster
galaxy redshift survey (Bardelli et al. 1997), and of the implied peculiar 
velocities, will help in discriminating among these two scenarios: in fact, 
peculiar velocities of the order of $1000$ km/s at the center of the Shapley 
Concentration would point toward the hypothesis of a cluster-cluster 
collision, while significantly lower velocities would be consistent with the
second scenario.

%

%
\end{document}